# Thermonuclear Dynamo inside an Alfvén Black Hole

F. Winterberg

*University of Nevada, Reno*

## Abstract

As in an acoustic black hole where the fluid is moving faster than the speed of sound and where the sound waves are swept along, in an Alfvén black hole the plasma is moving faster than the Alfvén velocity, with the Alfvén waves swept along and eliminated as the cause of the magneto hydrodynamic instabilities. To realize an Alfvén black hole, it is proposed to bring a plasma into rapid rotation by radially arranged lumped parameter transmission lines intersecting the plasma under an oblique angle. The rotating plasma slides frictionless over magnetic mirror fields directed towards the rotating plasma, with the mirror fields generated by magnetic solenoids positioned at the end of each transmission line. It is then shown that, with this configuration one can realize a thermonuclear dynamo, which also can serve as the analogue of a magnetar.

## 1. Introduction

The concept of an acoustic black hole has emerged from analogies between general relativity and frictionless compressible fluid dynamics. These analogies permit to obtain an acoustic metric where the velocity of light is replaced by the velocity of sound [1], and where in an acoustic black hole the event horizon is replaced by the surface separating the subsonic from the supersonic region [2,3]. Disturbances launched from the subsonic region are totally reflected from the surface of separation. This effect has its counterpart in optics, where a light wave propagating with the velocity $c/n$, ($n > 1$ refractive index) is totally reflected for angles of incidence $\sin\alpha > 1/n$, entering vacuum where the velocity is $c > c/n$.

Inside an acoustic black hole, where the fluid velocity exceeds the velocity of sound, sound waves entering this region are swept along. Likewise, in an Alfvén black hole, where the plasma velocity exceeds the Alfvén velocity, Alfvén waves are swept along. With the Alfvén waves swept along, the magnetohydrodynamic instabilities are eliminated.

In gravitationally confined plasmas, the gravitational forces normally exceed the magnetohydrodynamic forces, (even in the extreme case of a magnetar, a neutron star with an extremely large magnetic field). There the magnetohydrodynamic instabilities have only an "ornamental" effect on an otherwise stable configuration. This suggests to replace in a laboratory plasma the gravitational forces with centrifugal forces, which are larger than the magnetohydrodynamic forces if the rotational velocity exceeds the Alfvén veloctiy, creating an Alfvén black hole. This idea though raises two problems:

1. How to accelerate the plasma to ultrafast super-Alfvén velocities, and

2. How to keep the plasma away from the external wall.

## 2. Accelerating the plasma to super-Alfvén rotational velocities

One way to bring a plasma into rapid rotation is by placing it in between two coaxial conductors immersed in an axial magnetic field. Applying a voltage between the inner and outer conductor leads to an azimuthal $\mathbf{E} \times \mathbf{B}$ drift motion at low plasma densities. At high plasma densities, it leads to rapidly rotating arcs, driven by the $\mathbf{I} \times \mathbf{B}$ body force, where **I** is the electric current flowing in the radial direction from the inner to the outer conductor. However, because the plasma can there not be kept away from the outer conductor, this leads to large friction losses preventing the plasma from reaching super-Alfvén velocities.

To overcome this problem, it is proposed to bring the plasma into rapid rotation by a traveling magnetic wave moving at a high speed along its outer surface, resp. equator, with the wave driven by a circular arrangement of lumped parameter transmission lines intersecting the plasma under an oblique angle, whereby the plasma slides at its outer surface over magnetic mirrors (see Fig. 2). For maximum impedance matching, the phase velocity of the traveling magnetic waves in the lumped parameter transmission lines must be equal to the velocity of the traveling magnetic wave moving around the plasma. Following its attainment of super-Alfvén rotational velocities, the energy supplied by the lumped parameter transmission lines must thereafter only overcome the residual friction of the imperfect gliding of the plasma over the magnetic mirrors.

The phase velocity of the lumped parameter transmission line in electrostatic cgs units is given by [4,5]:

$$U = c/\sqrt{\boldsymbol{LC}} \tag{1}$$

where $\boldsymbol{L}$ and $\boldsymbol{C}$ are the inductance and capacitance per unit length of the line.

Writing (1) as follows:

$$U = c/n \tag{2}$$

One can assign a refractive index $n = \sqrt{\boldsymbol{LC}}$ to a lumped parameter transmission line. For the choices $U = \mathrm{v}_A \sim 10^8\, cm/s$, one has $\sqrt{\boldsymbol{LC}} \approx 300$.

To have 10 Alfvén wave lengths $\lambda$ spread over the circumference $2\pi R$ of a plasma spherical in shape, one has $2\pi R/\lambda = 10$, or $kR = 10$. Setting $U = \mathrm{v}_A = \omega/k = 10^8\, cm/s$, one has $\omega = k\mathrm{v}_A = 10^9/R$. Taking the example $R = 50cm$, one finds that $\omega = 2\times10^7 s^{-1}$, or $\nu = \omega/2\pi \approx 3\times10^6 s^{-1}$.

From Thomson's Formula:

$$\omega = c/\sqrt{LC} \tag{3}$$

where L and C are the inductance and capacitance of the transmission line given in cm, one obtains in conjunction with $U = \omega/k = c/\sqrt{\boldsymbol{LC}}$ that for each transmission line:

$$\sqrt{LC} = \sqrt{\boldsymbol{LC}}/k = \sqrt{\boldsymbol{LC}}R/10 = 15 \text{ meters} \tag{4}$$

The energy deposited into the plasma is in part kinetic energy of the rotational motion, and in part energy converted into heat. Assuming that the energy transfer from the transmission lines into the plasma is as a completely inelastic collision, ½ of the energy will go into the rotational kinetic energy with the other ½ going into heat. If the rotating plasma slides without friction along the outer wall, it retains its angular momentum and its rotational energy, while the heat is radiated away. With the radiative heat losses compensated by thermonuclear reactions, a steady state is established.

A spherically shaped plasma, with a radius $R$, and a number density $n$, requires for ignition the energy

$$E = (4\pi/3)R^3 \times 3nkT \tag{5}$$

where for the deuterium-tritium reaction $kT \approx 10^{-8}$ erg ($T \approx 10^8 {}^\circ K$). Assuming that $n \approx 10^{16}$ cm$^{-3}$ and $R$ = 50 cm, one finds that $E \approx 10^{14}$ erg = $10^7$ J. This energy has to be deposited in a time shorter than the bremsstrahlung loss time $\tau_R$, which at $n = 10^{16}$ cm$^{-3}$ and $T = 10^8 {}^\circ K$ is $\tau_R \approx 0.1$ s. With the assumed completely inelastic energy transfer, twice as much energy, i.e. $E = 2 \times 10^7$ J, has to be deposited into the plasma in the time $\tau_R = 0.1$ s, at a power of $P = 2 \times 10^8$ W. If there are as many transmission lines as there are Alfvén wave lengths over the circumference $2\pi R \approx 3$ meter, for example (10 Alfven wave lengths, each 30 cm long), the power of each lumped parameter transmission line is $2 \times 10^7$ W.

## 3. Stability

With the magnetic mirror field $B_0 >> B$, and $\rho \mathrm{v}^2 \sim B_0 / 4\pi$, the surface of the plasma is stable because the mirror field is convex towards the plasma. But for $\mathrm{v} >> \mathrm{v}_A = B/\sqrt{4\pi\rho}$, the plasma is also stable with regard to magnetohydrodynamic disturbances. This can be seen from the magnetohydrodynamic equation of motion (p plasma pressure, **j** electric current density):

$$\frac{\partial \mathbf{v}}{\partial \mathrm{t}} = -\frac{1}{\rho}\nabla \mathrm{p} - \frac{1}{2}\nabla \mathrm{v}^2 + \mathbf{v} \times \mathrm{curl}\,\mathbf{v} + \frac{1}{\rho \mathrm{c}}\mathbf{j} \times \mathbf{B} \tag{6}$$

For a uniformly rotating plasma one has

$$\mathbf{v} = \boldsymbol{\omega} \times \mathbf{r} \tag{7}$$

where **ω** is directed along the axis of rotation. Inserting (7) into (6) one has

$$\frac{\partial \mathbf{v}}{\partial t} = -\frac{1}{\rho}\nabla p - \omega^2 \mathbf{r} - 2\,\boldsymbol{\omega}\times\mathbf{v} + \frac{1}{\rho c}\mathbf{j}\times\mathbf{B} \tag{8}$$

The magnetohydrodynamic instabilities come from the last term on the r.h.s. of (8). For $v >> v_A$, where $(1/2)\rho v^2 >> B^2/8\pi$, the magnetic field lines are forced to align themselves with the lines of the super-Alfvén plasma flow. Then, if likewise **j** aligns itself with **ω**, one has $\mathbf{j} \perp \mathbf{B}$, since $\boldsymbol{\omega} \perp \mathbf{v}$, and $\boldsymbol{\omega} \perp \mathbf{B}$. With **ω** = (1/2)curl**v** and **j** = $(c/4\pi)$curl**B**, the magnetic forces are overwhelmed by the fluid flow forces if

$$|\mathbf{v}\times \mathrm{curl}\mathbf{v}| >> \left|\frac{\mathbf{B}\times \mathrm{curl}\mathbf{B}}{4\pi\rho}\right| \tag{9}$$

or if **v** = const**B** simply if

$$v >> v_A \tag{10}$$

A uniformly rotating fluid can also be viewed as a lattice of potential vortices. Here too, it is easy to understand why a magnetized plasma is stable for $v > v_A$, where v is the fluid velocity of the potential vortex. To prove stability we compare a linear pinch discharge where outside the pinch column curl**B** = 0, with a potential vortex where outside the vortex core curl**v** = 0 (see Fig. 4). Because of curl**B** = 0, the magnetic field strength gets larger with decreasing distance from the center of curvature of the magnetic field lines. For curl**v** = 0, the same is true for the velocity of a potential vortex. But whereas in the pinch discharge a larger magnetic field means a larger magnetic pressure, a larger fluid velocity means a smaller pressure by virtue of Bernoulli's theorem. Therefore, whereas a pinch column is unstable with regard to its bending, the opposite is true for a line vortex. Thus, a pinch column can be stabilized by placing it into a vortex provided $v >> v_A$ What is true for the m = 1 kink instability is also true for the m = 0 sausage instability by the conservation of circulation

$$Z = \oint \mathbf{v} d\mathbf{r} = \text{const.} \tag{11}$$

which implies that upon its pinching inside a vortex, $v \propto 1/r$, with $\rho v^2 \propto 1/r^2$, rising in the same proportion as the magnetic pressure, stabilizing the pinch against the m = 0 instability for $v > v_A$. And due to the centrifugal force, the vortex also stabilizes the plasma against the Rayleigh-Taylor instability.

With the magnetic field in the azimuthal and the current in the axial direction, the configuration can be seen as a large, vortex stabilized z-pinch.

The presence of two (or more) line vortices with the same sense of rotation and resulting from the vortex breakdown of the uniformly rotating fluid can not lead to an instability, because the interaction of two such line vortices leads to a circular motion of the vortices around each other [6].

## 4. The thermal convection pattern and dynamo action

It was shown by Cowling [7], that a magnetic field symmetric about an axis of rotation can not be maintained by a rotational symmetric motion, suggesting that the convective flow pattern must be twisted. In the geodynamo, twisting is caused by the Coriolis force, which plays a decisive role in establishing a dynamo by thermal convection.

Thermal convection has its cause in the buoyancy force, which here acts on the plasma confined in the flux tubes. In the presence of the radially outwards directed centrifugal force, the buoyancy force is radially directed inwards. In the rapidly rotating configuration, the thermonuclear reactions take place in a flux tube wound around the equator as shown in Fig. 3. It is from here that the convection pattern originates.

A magnetic field amplifying flow pattern similar to the one suggested by Parker [8] is here proposed with the sequence of events shown in Fig. 5. In this pattern, the flux tubes are folded and stretched. The convection directed towards the axis of rotation radially compresses a flux tube. As a result, a segment of the flux tube develops a dent, which by the Coriolis force is pushed sidewise, assuming a snake-like shape. The formation of the two sharp curves in the snake-like shape leads to an increase in magnetic energy, as can be seen as follows. With $r_o$ the radius of curvature of the flux tube, and with $B_o$ the magnetic field strength in its center, one has inside the tube (see Fig. 6)

$$B = B_o(r_o / r) \tag{12}$$

and for the magnetic energy density

$$\varepsilon_M = \frac{B^2}{8\pi} = \frac{B_o^2}{8\pi}\left(\frac{r_o}{r}\right)^2 \tag{13}$$

Integrating (13) over a segment with the angle $\varphi$, and from $r_o - \Delta r$, to $r_o + \Delta r$, where $\Delta r$ is the radius of the tube, one has for the magnetic energy per unit height of this segment

$$E_M = \frac{B_o^2 r_o^2}{8\pi}\int_0^{\varphi} d\varphi \int_{r_o-\Delta r}^{r_o+\Delta r} \frac{dr}{r} = \frac{B_o^2 r_o^2 \varphi}{8\pi}\ln\left(\frac{1+\Delta r / r_o}{1-\Delta r / r_o}\right) \tag{14}$$

Dividing (14) by $r_0\varphi$, one obtains the magnetic energy per unit length of the flux tube

$$\mathbf{E_M} = \frac{B_o^2 r_o}{8\pi}\ln\left(\frac{1+\Delta r / r_o}{1-\Delta r / r_o}\right) \tag{15}$$

or for $\Delta r << r_o$

$$\mathbf{E_M} \cong \frac{B_o^2}{4\pi}\left(\Delta r + \frac{1}{3}\frac{(\Delta r)^3}{r_o^2} + ...\right) \tag{16}$$

The increase in magnetic energy leads to a force per unit flux tube height and length

$$\mathbf{F} = -\frac{\partial \mathbf{E_M}}{\partial \mathbf{r_o}} = \frac{B_o^2}{6\pi}\frac{(\Delta r)^3}{\mathbf{r}_o^3} \tag{17}$$

vanishing for $r_o \to \infty$. This force tries to increase the radius of curvature $r_o$, which it can do by flipping over the snake-like pattern into a loop as shown in Fig. 5d, eventually leading to Fig. 5e, doubling the magnetic field. Together with the flux tubes containing hot plasma, which are moving towards the axis of rotation, there must be an equal number of flux tubes containing cooler plasma moving in the opposite direction.

For the high electrical conductivity of a thermonuclear plasma, the magnetic Reynolds number is much larger than one, as it is required for dynamo action.

## 5. The thermal convection velocity

In computing the thermal convection velocity we can apply the methods developed for the theory of the structure of stars [9]. Defining $\Delta\nabla T$ as the excess of the actual temperature gradient – in absolute amount – over the adiabatic temperature gradient, the convective heat flux per unit area is given by

$$H = c_p \rho \mathrm{v} \ell \Delta\nabla T \tag{18}$$

where $c_p$ is the specific heat at constant pressure, $\rho$ the density, v the convection velocity and $\ell$ the mixing length of the rising plasma. For stars one can set $\ell = R/10$, where $R$ is the radius of a star. In our case $R$ has to be set equal to the radius of the spheromak, with $l < R$, and we here too assume that $l = R/10$.

By order of magnitude one further has

$$(4\pi/3)\ell^3 q = 4\pi\ell^2 H \tag{19}$$

or

$$H = (1/3)\,\ell q \tag{20}$$

where $q$ is the heat released into the plasma per unit volume by the thermonuclear reactions. From (18) and (19) one obtains

$$\Delta\nabla T = \frac{q}{3\,c_p \rho\,\mathrm{v}}. \tag{21}$$

A further relation in the theory for the structure of stars is

$$\tfrac{1}{2}\rho\,\mathrm{v}^2 = \frac{\rho}{T}\Delta\nabla T \cdot \ell\int_0^\ell g(r)dr \tag{22}$$

where $g$ is the gravitational acceleration. In our case $g = r\omega^2$, where $\omega = \mathrm{v}/r$ is the vorticity, constant for uniform rotation, and one obtains from (22)

$$\mathrm{v}^2 = \frac{1}{T}\Delta\nabla T\,\omega^2\ell^3 \tag{23}$$

and with the help of (21)

$$\mathrm{v} = \ell\left[\left(q\,\omega^2\right)/\left(3\rho\,c_p T\right)\right]^{1/3}. \tag{24}$$

With $\rho c_\mathrm{v} = 3nkT$, $c_p / c_\mathrm{v} = \gamma$, where for hydrogen $\gamma = 5/3$, one has $\rho c_p = 5nkT$ and hence

$$\mathrm{v} = \ell\left[\left(q\omega^2\right)/\left(15nkT\right)\right]^{1/3}. \tag{25}$$

For $q$ we have to set for the DT reaction

$$\mathrm{q} = (1/4)n^2\langle\sigma\,\mathrm{v}\rangle\varepsilon_\alpha \tag{26}$$

where $\varepsilon_\alpha = 3.6\,\text{MeV} = 5.8\times10^{-6}$ erg is the energy set free into the He4 particles by the DT reaction, and where $\langle\sigma\text{v}\rangle$ is the nuclear reaction cross section velocity product averaged over a Maxwellian. At $T\sim10^8$ °K (10 keV) one has $\langle\sigma\text{v}\rangle\cong10^{-15}$ cm$^3$/s. We take the following example: $R = 50^2$ cm, $\ell = R/10 = 5$ cm, $R\omega = 10^8$ cm/s, $\omega = 2\times10^6$ s$^{-1}$, $n = 10^{16}$ cm$^{-3}$, $kT = 10^{-8}$ erg. We find $q = 1.4\times10^{11}$ erg/cm$^3$s, $\text{v}\cong4\times10^5$ cm/s, and $H = 2.3\times10^{11}$ erg/cm$^2$s.

Convection only occurs if the energy transport by heat conduction perpendicular to B is less than by convection. With $\nabla\text{T}$ the temperature gradient in absolute amount the conductive heat flux per unit area is given by

$$J_\perp = \kappa_\perp \nabla T \tag{27}$$

where for a fully developed ionized DT plasma [10]

$$\kappa_\perp = 2.34\times10^{-17}\frac{n^2\ln\Lambda}{\sqrt{T}B^2}\frac{erg}{s^0Kcm} \tag{28}$$

is the heat conduction coefficient for heat flow perpendicular to B. For the example $n = 10^{16}$ cm$^{-3}$, T = $10^8$ °K, $B^2/8\pi = nkT$, $\ln\Lambda\sim10$, one finds that $\kappa_\perp\sim10^3$ [erg/s°Kcm]. For an estimate of the temperature gradient we set $\nabla T\sim0.1T/l = 2\times10^6$ °K/cm, by which we obtain $J_\perp\sim2\times10^9$ erg/cm$^2$s, clearly less than H by two orders of magnitude.

The heat conduction $J_\parallel$ parallel to the equator of the rotating plasma is, with $\kappa_\parallel >> \kappa_\perp$, much larger and keeps the temperature in the azimuthal direction constant along the burn zone shown in Fig. 3.

To determine if the convection is laminar or turbulent, we have to determine the Reynolds number

$$\mathrm{Re} = \frac{\ell \mathrm{v}}{\nu_\perp} \tag{29}$$

where [10]

$$\nu_\perp = 2.5 \times 10^{-2} \frac{n \ln \Lambda}{\sqrt{T} B^2} \frac{cm^2}{s} \tag{30}$$

is the perpendicular kinematic viscosity for the mutual friction of the rising and falling plasma inside the magnetic flux tubes (see Fig. 5). For $B^2 / 8\pi = nkT$ and $\ln \Lambda \sim 10$, one finds that $\nu_\perp \sim 10^{14} T^{-3/2}$ [cm$^2$/s], and for T=$10^8$ °K that $\nu_\perp \sim 10^2$ cm/s. Hence $\mathrm{Re} \sim 5 \times 10^3$, just below the critical Reynolds number $\mathrm{Re}_{crit} \approx 10^{14}$, where turbulence sets in. With $\beta = (B^2 / 8\pi) / nkT < 1$ inside the plasma, $\nu_\perp$ is there larger and with it the Reynolds number smaller, making a laminar convection flow likely.

**6. Analogue of magnetar**

While a laboratory analogue of the geodynamo can be made with a rapidly rotating liquid metal, the concept presented here can be seen as the analogue of a magnetar, a neutron star with an extremely strong magnetic field generated by a convectively driven dynamo of hot nuclear matter, drawing its energy from nuclear reactions.

The mass of a magnetar is of the order of a solar mass, $M = 2 \times 10^{33}$ g, with a radius $R \approx 10^6$ cm. The gravitational acceleration at its surface is (G Newton's constant)

$$|\mathbf{g}| = \frac{GM}{R^2} \cong 10^{14} \text{ cm/s}^2 \tag{31}$$

Comparing it with the centrifugal acceleration of the rotating plasma at a speed of v = $10^8$ cm/s at its equator with the radius R = 50 cm, we find

$$\frac{v^2}{R} = 2\times10^{14}\ \text{cm/s}^2 \quad , \tag{32}$$

of the same order of magnitude.

Next we turn to the Alfvén velocity in a magnetar

$$v_A = \frac{B}{\sqrt{4\pi\rho}} \tag{33}$$

where $\rho = M/(4\pi/3)R^3 \cong 5\times10^{14}$ g/cm$^3$ is the density of a magnetar and $B = 10^{15} G$. We find that $v_A \cong 10^7$ cm/s, off by one order of magnitude from the analogue where $v_A \cong 10^7$ cm/s.

Following Hund [11], we can take the analogy one step further. With **F** the gravitational field vector, Newton's law states that

$$\text{div}\mathbf{F} = -4\pi G\rho \tag{34}$$

On the other hand, for a uniform rotation the centrifugal force vector is given by

$$\mathbf{F} = \omega^2\, \mathbf{r} \tag{35}$$

for which

$$\text{div}\mathbf{F} = 2\,\omega^2 \tag{36}$$

To make the connection to the analogue we set

$$-4\pi G\rho = 2\,\omega^2 \tag{37}$$

whereby

$$\rho = -\frac{\omega^2}{2\pi G} \tag{38}$$

is a negative repulsive mass density of the centrifugal force field, as the analogue of the attractive positive mass density of a magnetar. For $\omega = v_A/R = 2\times10^6\ \text{s}^{-1}$, we obtain $\rho \approx -10^{-19}$ g/cm$^3$, in absolute amount 4-5 orders of magnitude larger than the density of a magnetar. This negative mass density is not fictitious but rather presents a physical reality, as can be seen as follows:

The gravitational field **g** given by (31) has a negative energy density

$$\varepsilon_g = -\frac{\mathbf{g}^2}{8\pi G} \tag{39}$$

very much as the electric field of a charge e

$$E = \frac{e}{r^2} \tag{40}$$

has a positive electric energy density

$$\varepsilon_e = \frac{E^2}{8\pi} \tag{41}$$

But according to Einstein's equation E = mc$^2$, an energy density has a mass density. For the gravitational field of a magnetar, with the gravitational field given by (31), this mass density is

$$\frac{\mathbf{g}^2}{8\pi G c^2} \cong -7\times 10^{12}\ \text{g/cm}^3 \tag{42}$$

in absolute amount by about two orders of magnitude less than the mass density of the magnetar. On the other hand, the mass density of the centrifugal force field $\mathbf{F} = \omega^2\mathbf{r}$, for r = R = 50 cm, $\omega = 2\times 10^6$ s$^{-1}$, is

$$\frac{\mathbf{F}^2}{8\pi G c^2} = -3\times 10^{14}\ \text{g/cm}^3 \tag{43}$$

in absolute amount about 4 – 5 orders of magnitude less than the mass density $\rho \cong$ -10$^{19}$ g/cm$^3$, given by (38). But in a rotating reference frame there is, besides the centrifugal force field a Coriolis force field **C**

$$\mathbf{C} = 2c\boldsymbol{\omega} \tag{44}$$

with the equation of motion for a test particle in the rotating reference frame given by

$$\ddot{\mathbf{r}} = \mathbf{F} + \frac{\mathbf{v}}{c} \times \mathbf{C} \tag{45}$$

Computing the mass density of the Coriolis force field one obtains

$$\frac{\mathbf{C}^2}{8\pi G c^2} = -\frac{\omega^2}{2\pi G} \tag{46}$$

the same as in (38). Therefore, it is the negative mass density of the Coriolis field which is the source for the repulsive centrifugal force.

According to Mach's principle, it is the influence of the rotating celestial background which generates the Coriolis force field in the rotating reference frame. But because the Coriolis and centrifugal force fields are immediately felt by making a transformation from an inertial to a rotating reference system, the validity of Mach's principle must mean that it is the relative rotation of the zero point vacuum energy, not the distant stars brought into rotation around a rotating reference frame, which causes the Coriolis and centrifugal force fields [12].

As an analogue of a magnetar the large attractive gravitational field is in the proposed thermonuclear dynamo replaced by a large repulsive gravitational field of the vacuum in the rotating reference frame, very much as in the rotating liquid metal analogue of the geodynamo. The analogy between a magnetar and the proposed thermonuclear dynamo is significant, because we know from astronomical observations that a magnetar as a working thermonuclear dynamo is realized in nature. This should give us hope that its proposed analogue might work as well.

## 7. Final Comments

Apart from flowing at super-Alfvén velocities, the proposed configuration is similar to the analog of a geodynamo, realized with a rapidly rotating liquid metal inside a container [13, 14, 15]. In the real geodynamo, the motion of the liquid core is turbulent with a very large Reynolds number, but even there the earth still has a hydromagnetic dynamo. The establishment of a dynamo effect under these high turbulence, could not be theoretically confirmed until the arrival of supercomputers. Therefore, even if in the proposed configuration the motion is turbulent, dynamo action may still be possible.

The problem of the energy loss by micro-instabilities is here no problem, unlike in conventional magnetic plasma confinement configurations, where the plasma is not wall stabilized and not confined by a convex magnetic field. This is here possible due to the super- Alfvén flow over the wall covered with magnetic mirrors. And in case the plasma is turbulent, it will have a non-turbulent laminar boundary layer near the wall where the heat conduction is classical. Furthermore, it can be shown that Nernst currents are set up near the wall, keeping the plasma away from the wall, which otherwise would lead to large radiation energy losses in its vicinity [16].

## 8. Conclusion

1. With only 20% of the DT fusion energy going into the He4 particles, the total fusion energy output is 5 times larger than $q = 1.4\times10^{11}$ erg/cm$^3$s, that is $q_{tot} = 7\times10^{11}$ erg/cm$^3$s, or for a volume $V = \ell^3 = 10^3$ cm$^3$, equal to ~ $10^{15}$ erg/s = 100 MW.
2. With $\omega = 2\times10^6$ s$^{-1}$, and $R\omega \cong 10^8$ cm/s, the vorticity $\omega$ keeps the plasma stiff as long as

$$\omega > v_A / R \tag{41}$$

where $v_A$ is the Alfvén velocity. For a high plasma β-value, $v_A \cong v^i = 10^8$ cm/s, which means the plasma is stable for $\omega > 2 \times 10^6$ s$^{-1}$.

3. The idea is conceptually quite simple and could be tested in the laboratory by experiments similar to those used to test theories for the geodynamo, replacing the nuclear by a non-nuclear heat source.
4. The really novel feature of the concept is to bring a magnetized plasma into super Alfvén rotational speed by a magnetic traveling wave moving with a speed of ~ $10^8$ cm/s along its equator, not only keeping the plasma stable, but also acting as a hydromagnetic dynamo, highly desirable for a continuous operation.
5. As it was shown by R. F. Ellis, A. Case, R. Elton et al. [17], a rotating plasma is in addition also stabilized by shear flow, with the shear flow heating the plasma to thermonuclear temperatures by viscous dissipation.

It was the late H. Alfvén who once told the author that the one reason for the failure of hitherto proposed magnetic fusion devices might be that we had not yet found the right configuration.

## 10. Figure Captions

Figure 1. Generation of a traveling magnetic wave by megahertz transmitters.

Figure 2. Magnetic mirror and lumped parameter transmission line.

Figure 3. Thermonuclear burn zone in rapidly rotating plasma.

Figure 4. Pinch instability due to curl**B** = 0; Vortex stability due to curl**v** = 0 and the Bernoulli theorem.

Figure 5. Magnetic field amplification through thermal convection of magnetic flux tubes.

Figure 6. Computation of the magnetic energy in a curved segment of the flux tube.

**11. Figures**

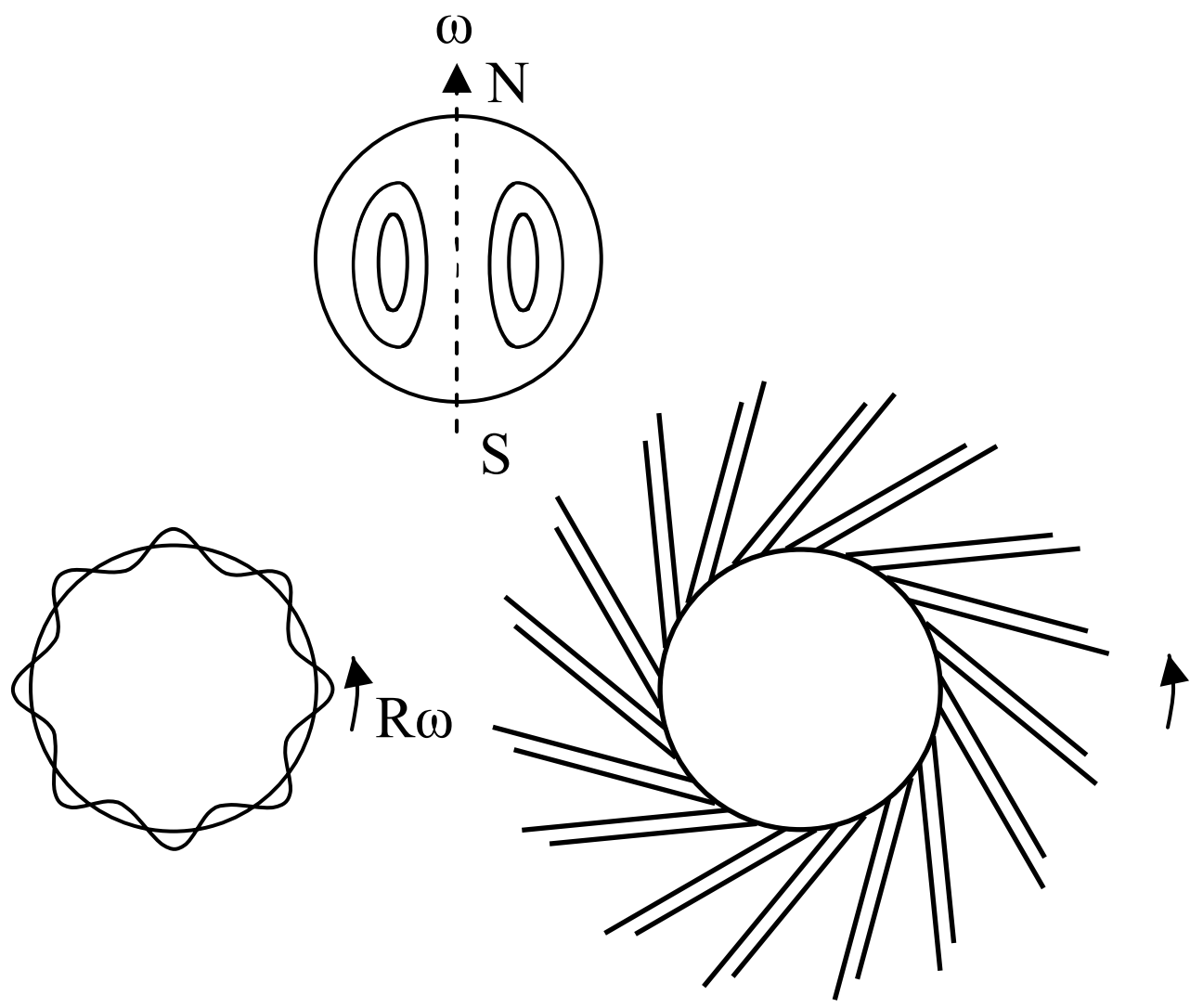

Figure 1

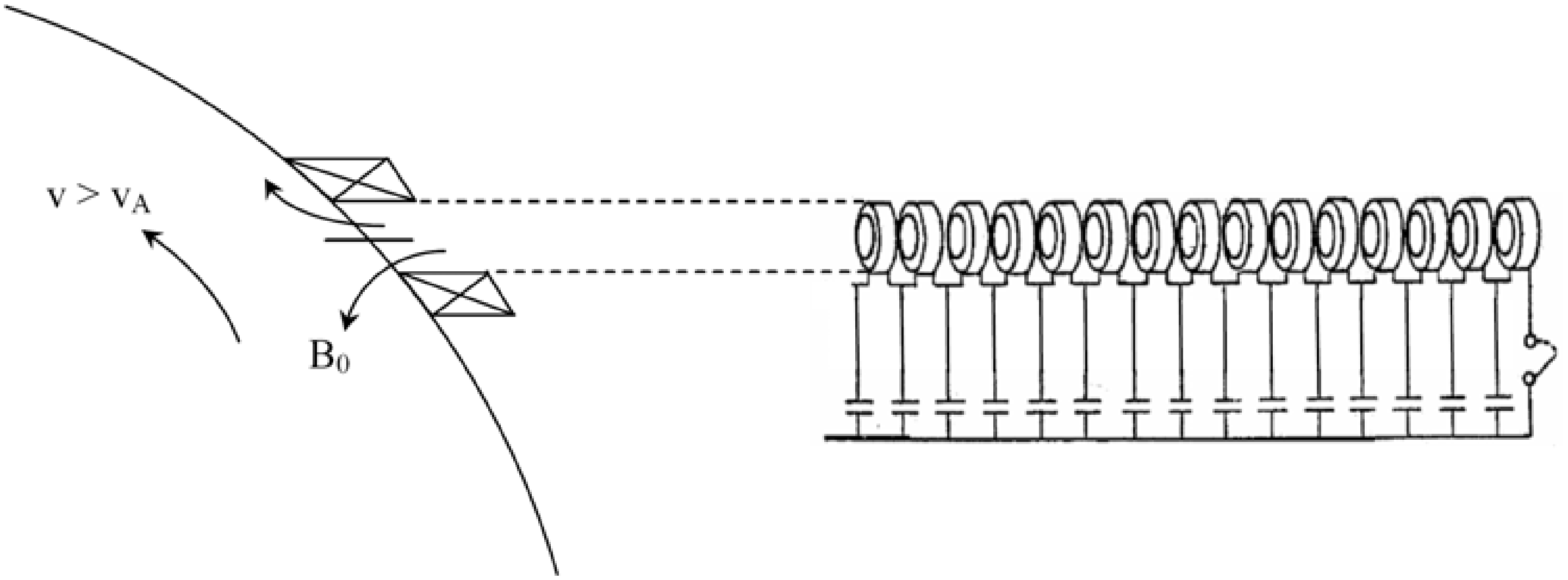

Figure 2

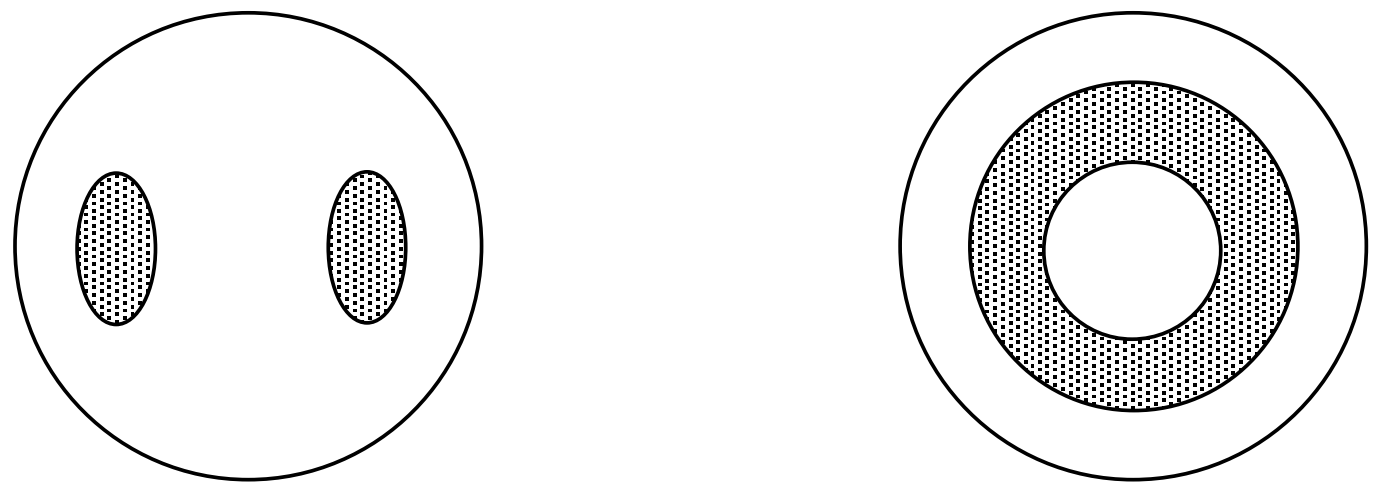

Figure 3

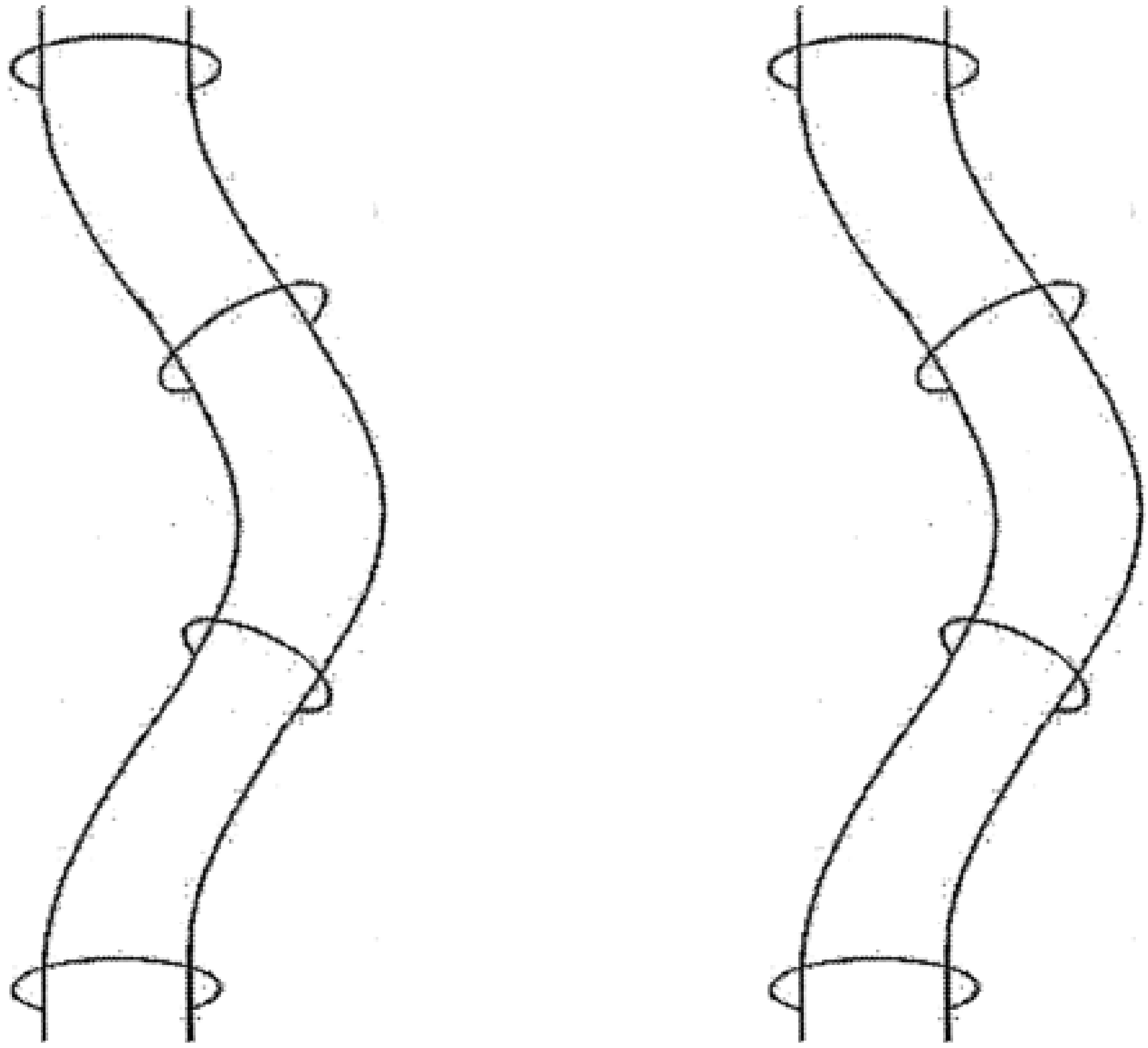

curl**B** = 0
unstable

curl**v** = 0
stable

Figure 4

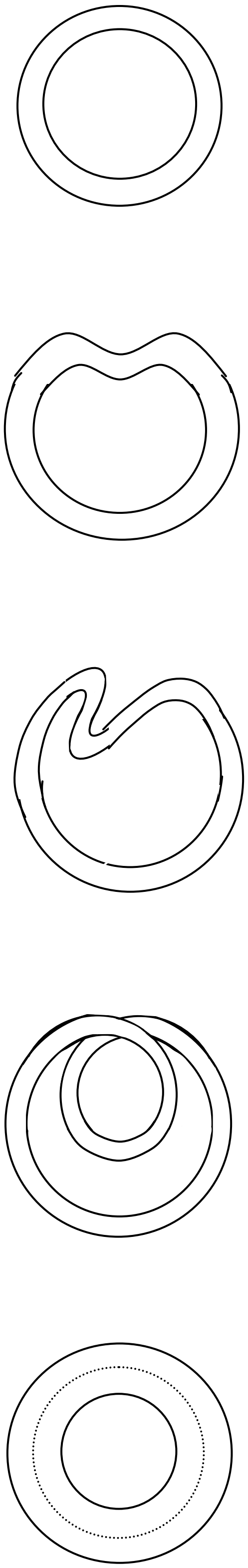

Figure 5

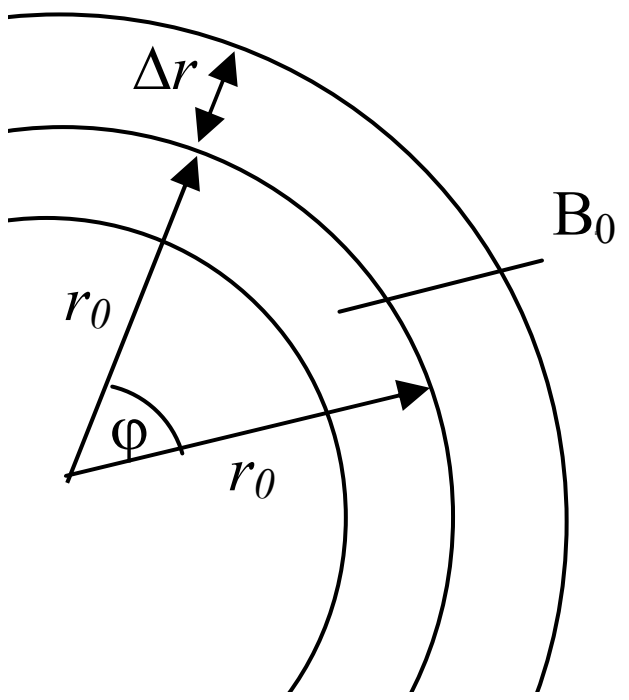

Figure 6